\documentclass[twocolumn,showpacs,floats,floatfix,superscriptaddress,aps,pra]{revtex4-1}
\usepackage{amsfonts}
\usepackage{amssymb}
\usepackage{amsmath}
\usepackage{calc}
\usepackage{graphicx}
\usepackage{bm}

\usepackage[normalem]{ulem}
\usepackage{amsmath,amssymb}
\usepackage{multirow}
\usepackage{xcolor,soul}
\usepackage{srcltx}
\usepackage{hyperref,graphicx}

\def\be{ \begin{equation}}
\def\ee{ \end{equation}}
\def\bea{ \begin{eqnarray}}
\def\eea{ \end{eqnarray}}
\def\bse{ \begin{subequations}}
\def\ese{ \end{subequations}}
\def\bc{ \begin{center}}
\def\ec{ \end{center}}

\begin{document}

\author{Stefano Longhi$^{*}$} 
\affiliation{Dipartimento di Fisica, Politecnico di Milano and Istituto di Fotonica e Nanotecnologie del Consiglio Nazionale delle Ricerche, Piazza L. da Vinci 32, I-20133 Milano, Italy}
\email{stefano.longhi@polimi.it}

\title{Oscillating potential well in complex plane and the adiabatic theorem}
  \normalsize






%
\bigskip
\begin{abstract}
\noindent  
A quantum particle in a slowly-changing potential well $V(x,t)=V(x-x_0(\epsilon t))$, periodically shaken in time at a slow frequency $\epsilon$, provides an important quantum mechanical system where the adiabatic theorem fails to predict the asymptotic dynamics over time scales longer than $ \sim 1 / \epsilon$.  Specifically, we consider a double-well potential $V(x)$ sustaining two bound states spaced in frequency by $\omega_0$ and periodically-shaken in complex plane. Two different spatial displacements $x_0(t)$ are assumed:  the real spatial displacement $x_0(\epsilon t)=A \sin (\epsilon t)$, corresponding to ordinary Hermitian shaking, and the complex one $x_0(\epsilon t)=A-A \exp( -i \epsilon t)$, corresponding to non-Hermitian shaking.
When the particle is initially prepared in the ground state of the potential well, breakdown of adiabatic evolution is found for both Hermitian and non-Hermitian shaking whenever the oscillation frequency $\epsilon$  is close to an odd-resonance of $\omega_0$. However, a different physical mechanism underlying nonadiabatic transitions is found in the two cases. 
For the Hermitian shaking, an avoided crossing of quasi-energies is observed at odd resonances and nonadiabatic transitions between the two bound states, resulting in Rabi flopping, can be explained as a multiphoton resonance process. For the complex oscillating potential well, breakdown of adiabaticity arises from the appearance of Floquet exceptional points at exact quasi energy crossing.
\end{abstract}



\maketitle

\section{Introduction}
The evolution of a quantum system under external adiabatic driving has been of fundamental interests to
physicists since the earlier days of quantum mechanics \cite{r1,r2}. A major result in quantum adiabatic evolution is provided by the 
the quantum adiabatic theorem (QAT) \cite{r1,r2,r3,r4}, which finds widespread applications in several areas of physics 
such as in  atomic and molecular physics \cite{r5,r6,r7}, quantum Hall physics \cite{r8}, the physics of geometric phase \cite{r9}, quantum computation \cite{r10,r11,r12}, quantum annealing \cite{r13,r14,r14bis} and quantum simulations \cite{r15} to mention a few. 
In its simplest form, as originally proposed by Born and Fock \cite{r1}, the QAT applies to a quantum system with discrete and non-degenerate energy levels and states that, if the  system is initially prepared in an instantaneous eigenstate (commonly the ground state) of the slowly-changing time-dependent
Hamiltonian $\hat{H}(\epsilon t)$, with an instantaneous eigenvalue $E_0=E_0(\epsilon t)$ which remains separated all the time by a finite gap from the rest of the spectrum, in the $\epsilon \rightarrow 0$ limit the system evolves remaining in the same instantaneous eigenstate, up to a multiplicative phase factor.
Several extensions of the QAT  theorem, that include the cases of a Hamiltonian with a continuous energy spectrum, gapless Hamiltonians, and time-periodic Hamiltonians with slowly-changing parameters, have been subsequently reported \cite{r2,r3,r4,r16,r17,r18}.\par
 While the correctness of the QAT is beyond any dispute, some inconsistencies have been disclosed when attempting to apply the QAT to certain Hamiltonian models \cite{r19,r20}. The origin and explanation of such inconsistencies have raised a rather lively debate among physicists over the past decade, and several facets of the problem have been discussed sometimes with different views \cite{r21,r22,r23,r23bis,r24,r25,r26,r27,r28,r29,r30,r31,r32,r33,r34,r35}.  Rather generally, failure of adiabatic following is observed when the Hamiltonian varies on different time scales, or in case the evolution of the quantum state is observed at extremely long time scales and the Hamiltonian contains oscillating terms \cite{
r23,r23bis,r26,r27,r31,r34,r35}. Indeed, the QAT ensures adiabatic following provided that the Hamiltonian $\hat{H}$ changes with time as $\hat{H}(x,\epsilon t)$, where $\epsilon$ is assumed small, and the time dependence vanishes after some finite time, that is,
$\hat{H}(x,\epsilon t) = \hat{H}(x,\infty)$ for $t>t_{\infty}$, typically $t_{\infty}$ of order $\sim 1/ \epsilon$ \cite{r34}.
When the slow change never really stops or continues for a time much longer than $1/ \epsilon$,
the predictions of the adiabatic theorem can fail. This happens, for example, when the Hamiltonian undergoes a periodic change (though small and at extremely low frequency $\epsilon$) and the evolution of the system is observed for an extremely long time: after many oscillation cycles, for special driving frequencies  corrections to the adiabatic solution can sum up  constructively, resulting in nonadiabatic transitions and Rabi flopping between energy levels \cite{r35}. Such nonadiabatic transitions show similar features to field-induced multiphoton resonances  and multiphoton Rabi oscillations encountered in laser-driven atomic systems \cite{r36,r37,r38}.\par
Recently, great attention has been devoted to extend the conditions of the adiabatic theorem to non-Hermitian Hamiltonians \cite{r39,r40,r41,r42,r43,r44,r45,r46,r47,r48}. In non-Hermitian systems, the usual approximations and criteria of the QAT are not necessarily valid, and several results have been found concerning extensions and breakdown of the adiabatic theorem \cite{r40,r41,r42,r44,r45}.  A unique feature of non-Hermitian Hamiltonians, as compared to Hermitian ones, is the appearance of exceptional points (EPs), i.e. spectral singularities in the point spectrum of the Hamiltonian corresponding to the coalescence of two (or more) eigenvalues and of corresponding eigenfunctions \cite{r49,r50,r51,r51bis,r51tris}.  Interestingly, EPs can deeply modify adiabatic evolution, with the appearance of a chiral behavior when the Hamiltonian is slowly varied to encircle an EP: while adiabatic following is observed when the EP is encircled in one direction (e.g. clockwise), nonadiabatic transitions are observed when the EP is encircled in the opposite direction (e.g. counter-clockwise) \cite{r42,r43,r47,r48}. Recent experimental progress in engineered electromagnetic, electronic and optical systems has made it possible to access the intriguing properties of non-Hermitian Hamiltonian models and the impact of non-Hermitian dynamics on adiabatic evolution in an unprecedented way. For example, the chiral behavior of EPs has been recently demonstrated in a classical system using deformed metallic waveguides \cite{r48}.  \par
In this work we show that periodic shaking of a potential well in 'complex' space provides a noteworthy example where breakdown of the adiabatic theorem can be observed in Hermitian and non-Hermitian realms under different physical mechanisms. The periodically-shaken double-well potential has been widely investigated in the open literature as a basic model of tunneling control in different areas of physics \cite{r52}.  Depending on the strength and frequency of the shaking, suppression
or enhancement of tunneling can be observed \cite{r53,r54,r55,r56}. Here we consider a double-well potential $V(x)$, sustaining two bound states spaced in frequency by $\omega_0$, which is periodically-shaken in 'complex' plane leading to a time-dependent potential $V(x,t)=V(x-x_0(t))$. The main reason of considering a 'complex'  shaking of the potential, in addition to a real one, is to reveal a novel mechanism of failure of the adiabatic theorem which is peculiar to non-Hermitian potentials and related to the appearance of Floquet EPs. While in the oscillating Hermitian potential failure of adiabatic theorem results in a kind of Rabi flopping, in the oscillating non-Hermitian potential failure of the adiabatic theorem results in the emergence of a  dominant state and a chiral dynamical behavior, which is impossible to observe in the Hermitian case.
We assume either a real spatial displacement (Hermitian shaking), $x_0(\epsilon t)=A \sin (\epsilon t)$, or a complex spatial displacement (non-Hermitian shaking), $x_0(\epsilon t)=A-A \exp(- i \epsilon t)$. 
In the former case  the potential $V(x,t)$ remains real and shape invariant, whereas in the latter case the potential becomes complex and it is not anymore shape invariant. By application of rigorous Floquet theory \cite{r57}, we show that breakdown of adiabatic following is observed for both Hermitian and non-Hermitian periodic shaking when the driving frequency $\epsilon$ is tuned close to the critical frequencies $\epsilon_N$ satisfying the odd-resonance condition $\epsilon_N \simeq \omega_0 / (2N-1)$ ($N=1,2,3,...$). However, the physical mechanism underlying nonadiabatic transitions is very distinct in the two cases. For the Hermitian shaking, nonadiabatic transitions arise from a multiphoton resonance process near avoided crossings of quasi energies and lead to Rabi flopping between the two levels,  with a mechanics similar to the one recently investigated in Ref.\cite{r35}. On the other hand, for the complex oscillating potential well breakdown of the adiabatic theorem is rooted into the appearance of a Floquet EP, i.e. a singular regime where coalescence of both quasi energies and Floquet eigenstates occurs. 
 \begin{figure}[htbp]
  \includegraphics[width=87mm]{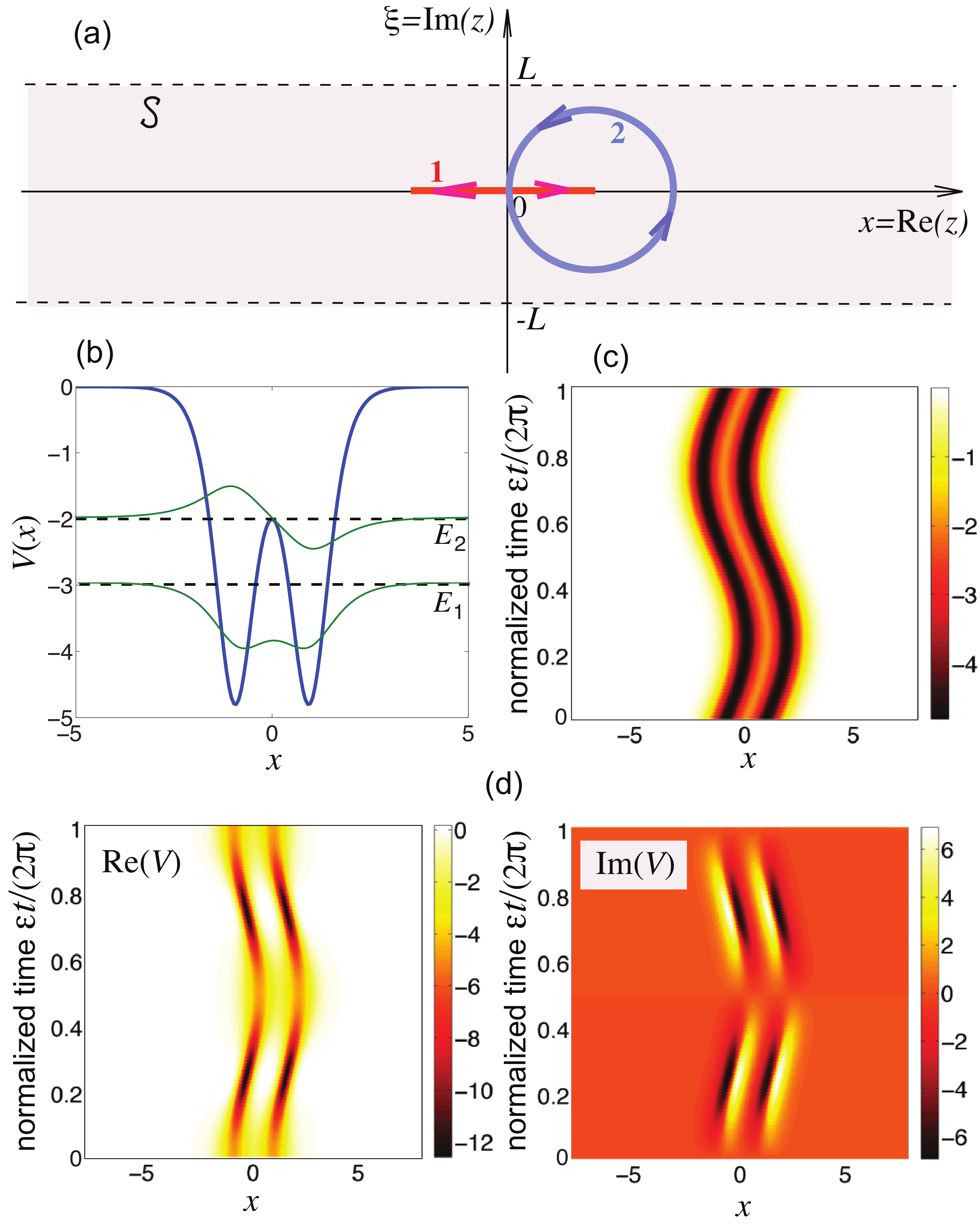}\\
   \caption{(color online) Periodical shaking of a potential well in complex plane. (a) Closed path $x_0=x_0(\epsilon t)$ of the oscillation in complex plane $z=x+i \xi$ . The potential $V$ is holomorphic in the stripe $\mathcal{S}: |\xi|<L$ of the complex plane. Path 1 on the real axis $x$ corresponds to the Hermitian shaking $x_0(\epsilon t)=A \sin(\epsilon t)$, whereas path 2 embedded in the stripe $\mathcal{S}$ corresponds to the non-Hermitian shaking $x_0(\epsilon t)=A-A \exp(-i \epsilon t)$. (b) Double-well potential sustaining two bound states with energies $E_1=-\sigma_1^2$ (ground state) and $E_2=-\sigma_2^2$ (excited state) with $\sigma_1=\sqrt{3}$ and $\sigma_2=\sqrt{2}$  [Eq.(30)]. The corresponding wave functions $u_1(x)$ and $u_2(x)$ are depicted by thin solid lines. (c-d) Behavior of the periodically-shaken potential well, over one oscillation cycle, for (c) Hermitian shaking $x_0(\epsilon t)=A \sin (\epsilon t)$ with $A=1$, and (d) non-Hermitian shaking $x_0(\epsilon t)=A-A \exp(-i \epsilon t)$ with $A=0.6$. In (c) the potential remains real, whereas in (d) the potential becomes complex. The two panels in (d) depict the real (left panel) and imaginary (right panel) parts of the potential. }
\end{figure}
\section{Periodically-shaken potential well in complex plane}
\subsection{Model and basic equations}
We consider the dynamics of a quantum particle in a slowly-shaken one-dimensional quantum well, which in scaled units is described by the dimensionless Schr\"odinger equation for the wave function $\psi=\psi(x,t)$
\begin{equation}
i \frac{\partial \psi}{\partial t}=-\frac{\partial^2 \psi}{\partial x^2}+V(x-x_0 (\epsilon t)) \psi \equiv \hat{H}(x, \epsilon t) \psi
\end{equation}
where $V(x)$ is the potential well at rest, $x_0=x_0( \epsilon t)$ is the time-dependent spatial displacement, and $x_0(0)=0$. The adiabatic limit corresponds to take $ \epsilon \rightarrow 0$ and to consider the dynamics for long times, namely up to the time scale of order $ \sim 1 / \epsilon$ or longer. For a periodically-shaken potential, $x_0(\epsilon t)$ is a periodic function of time with period $T_{\epsilon}= 2 \pi / \epsilon$. The potential $V(x)$ is a real function of space variable $x$, with $V(x) \rightarrow 0$ as $x \rightarrow \pm \infty$. We assume that $V(x)$ can be analytically prolonged into the complex plane $z=x+i \xi$ in a stripe $\mathcal{S}$: $|{\rm Im}(z)|=| \xi | < L$ embedding the real $x$ axis, with $|V(z)| \rightarrow 0$ as $|z| \rightarrow \infty$, $z\in \mathcal{S}$. The spatial displacement $x_0(\epsilon t)$ is generally assumed to be complex, describing a closed loop inside the stripe $\mathcal{S}$ of analyticity of $V(z)$ [Fig.1(a)]. We note that, for a real spatial displacement $x_0(\epsilon t)$ (Hermitian shaking), the potential $V(x,t)=V(x-x_0(t))$ is real and shape-invariant at any time $t$ [Fig.1(b) and (c)]: the particle dynamics corresponds to the ordinary Hermitian dynamics in a shape-invariant and periodically-shaken potential well \cite{r52}. For a complex spatial displacement $x_0(\epsilon t)$ (non-Hermitian shaking), the potential $V(x,t)$ is not anymore shape invariant and becomes a complex function [Fig.1(d)]: in this case the dynamics is described by a time-periodic non-Hermitian Hamiltonian $\hat{H}(\epsilon t)$. We assume that the potential well $V(x)$ sustains $N$ non-degenerate bound states $| u_1(x) \rangle$, $|u_2(x) \rangle$, ..., $|u_N(x) \rangle$ with energies $E_1<E_2<...<E_N <0$. Since the potential $V(x)$ is real, any eigenfunction $u_n(x)$ can be assumed to be real as well, and the orthonormality conditions
\begin{equation}
\int_{-\infty}^{\infty} dx \; u_n(x)u_m(x)= \delta_{n,m}
\end{equation} 
hold. The eigenfunctions $u_n(x)$ can be analytically prolonged in the stripe $\mathcal{S}$ of the complex $z$ plane, where they do not show poles neither branch cuts. Since $\hat{H}(x, \epsilon t)=H(x-x_0(\epsilon t),0)$, the instantaneous eigenfunctions of $\hat{H}(x, \epsilon t)$ at time $t$ are merely given by $|u_n(x-x_0(\epsilon t)) \rangle$ with energies $E_n(t)=E_n$ ($n=1,2,...,N$). This means that the instantaneous energies do not change in time, while the instantaneous eigenfunctions at time $t$ are simply obtained from the initial ones by application of the spatial displacement $x \rightarrow x-x_0(\epsilon t)$. Note that, for any given time $t$ the orthonormality conditions
\begin{equation}
\int_{-\infty}^{\infty} dx \; u_n(x-x_0(\epsilon t) )u_m(x-x_0(\epsilon t))= \delta_{n,m}
\end{equation}
hold. This follows from the fact that the integral on the left hand side of Eq.(3) can be computed by deformation of the contour path in complex plane inside the stripe $\mathcal{S}$ of analyticity, so as to coincide with the integral on the real $x$ axis [Eq.(2)]. Note that the integral on the left hand side of Eq.(3) is not the ordinary (Hermitian) scalar product of $|u_n(x-x_0(\epsilon t)) \rangle$ and $|u_m (x-x_0(\epsilon t)) \rangle$ when $x_0(\epsilon t)$ is complex, indicating that in the non-Hermitian case the eigenfunctions cease to be orthonormal under the ordinary Hermitian scalar product.\\ 
In the spirit of the adiabatic approximation and neglecting excitation into the continuum of states, we look for a solution to Eq.(1) of the form
\begin{eqnarray}
\psi(x,t)   & = & \sum_{n=1}^N c_n(t)  u_n\left( x-x_0(\epsilon t) \right)  \exp \left[ -i \int_0^t d\eta E_n(\eta)\right] \nonumber \\
& = & \sum_{n=1}^N c_n(t)  u_n \left( x-x_0(\epsilon t) \right)  \exp (-i E_n t)
\end{eqnarray}
with $c_n(0)=\delta_{n,1}$. The evolution equations of the complex amplitudes $c_n(t)$ are readily obtained after substitution of the Ansatz (4) into Eq.(1) and using the orthonormal conditions (3). One has
\begin{equation}
i \frac{dc_n}{dt}= \epsilon \dot{x}_0( \epsilon t) \sum_m \kappa_{n,m} c_m \exp [i(E_n-E_m)t]
\end{equation}
where we have set
\begin{equation}
\kappa_{n,m} \equiv i \int_{-\infty}^{\infty} dx \; u_n(x) \frac{du_m}{dx}.
\end{equation}
and where the dot denotes the derivative with respect to the argument $\epsilon t$ of $x_0$.
After integration by parts, from Eq.(6) it readily follows that the diagonal elements $\kappa_{n,n}$, which account for geometric (Berry) phase, vanish; whereas the off-diagonal elements are purely imaginary with $\kappa_{n,m}=\kappa^*_{m,n}$. For an Hermitian shaking ($x_0$ real), norm conservation implies $\sum_n |c_n(t)|^2=1$, however for the non-Hermitian shaking ($x_0$ complex) conservation of the norm is not ensured, and unbounded growth or decay of the amplitudes $c_n$ could be observed. In both cases, we say that the system undergoes adiabatic following provided that $|c_n(t)|^2 \ll |c_1(t)|^2$ for any $n \neq 1$ and for unbounded time $t$. 
Provided that the energy $E_1$ is spaced from the excited energy level $E_2$ by a sufficient gap, the QAT ensures that adiabatic following is met for a time scale at least of order $ \sim 1 / \epsilon$, i.e. at least for a few oscillation cycles of the shaking. However, from the QAT never can be said about the evolution of amplitudes $c_n$ at longer time scales, where failure of the adiabatic following could be observed.

\subsection{Two-level model}
Here we focus our analysis to the case of two bound states, i.e. we assume that the potential well sustains two bound states solely $u_1(x)$ (ground state) and $u_2(x)$ (excited state), with energies $E_1$ and $E_2$, or that excitation to higher excited states is negligible. An example of a potential well sustaining two bound states and periodically shaken in complex plane will be discussed in Sec.IV. We will also assume harmonic oscillation at frequency $\epsilon$ by assuming
\begin{equation}
x_0(\epsilon t)=-iA_1 \exp(i \epsilon t) +iA_2 \exp(-i \epsilon t)+i(A_1-A_2).
\end{equation}
Hermitian shaking is obtained by taking $A_1=A_2=A/2$, yielding $x_0(\epsilon t)=A \sin ( \epsilon t)$. The evolution equations for the amplitudes $c_1$ and $c_2$ [Eq.(5)] read
\begin{eqnarray}
i \frac{dc_1}{dt} & = &  -i \epsilon  \left[ V_1 \exp(i \epsilon t)+V_2 \exp(-i \epsilon t) \right] c_2 \exp(-i \omega_0 t)  \;\;\;  \\
i \frac{dc_2}{dt} & = & i \epsilon  \left[ V_1 \exp(i \epsilon t)+V_2 \exp(-i \epsilon t) \right] c_1 \exp(i \omega_0 t) 
\end{eqnarray}
where we have set
\begin{equation}
V_1 \equiv  \kappa A_1 \; , \;\;  V_2 \equiv  \kappa A_2 \; , \;\; \kappa \equiv \int_{-\infty}^{\infty} dx u_2(x) \frac{du_1}{dx}
\end{equation}
and $\omega_0 \equiv E_2-E_1$.
After setting
\begin{equation}
c_1(t)=a_1(t) \exp(-i \omega_0 t/2) \; ,\;\; c_2(t)=i a_2(t) \exp(i \omega_0 t/2)
\end{equation}
Eqs.(8) and (9) can be cast in the form
\begin{eqnarray}
i \frac{da_1}{dt} & = &  -\frac{\omega_0}{2}a_1 + \epsilon f(\epsilon t) a_2 \;\;\; \\
i \frac{da_2}{dt} & = & \frac{\omega_0}{2}a_2 +\epsilon  f(\epsilon t ) a_1 
\end{eqnarray}
where the modulation function $f(\epsilon t)$ is defined by
\begin{equation}
 f(\epsilon t) \equiv V_1 \exp(i \epsilon t)+V_2 \exp(-i \epsilon t).
\end{equation}

  \begin{figure}[htbp]
  \includegraphics[width=85mm]{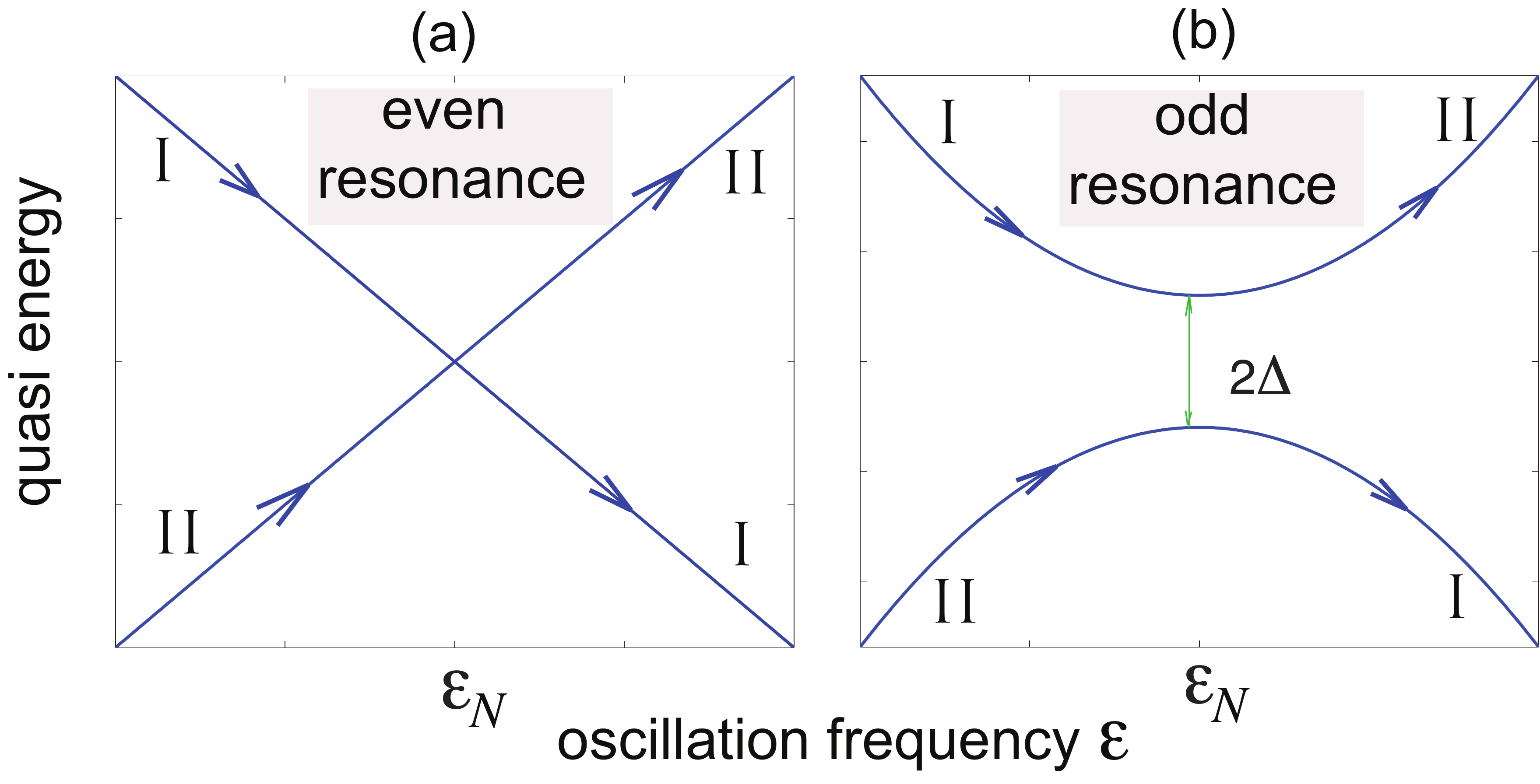}\\
   \caption{(color online) Schematic behavior of the quasi energies versus oscillation frequency $\epsilon$ near a resonance frequency $\epsilon_N \simeq \omega_0 /N$ for Hermitian shaking.
   I and II denote Floquet eigenstates with  dominant level-1 ($\mathbf{W}_1^{(WKB}$) and level-2 ($\mathbf{W}_2^{(WKB)}$), respectively.  
    In (a) the quasi energies cross at $\epsilon=\epsilon_N$, corresponding to Hermitian degeneracy. This case is observed for even resonances ($N$ even). As $\epsilon$ is continuously varied to cross the resonance, the Floquet eigenstates in each branch remain level-1 or level-2 dominant, as shown in the figure. Panel (b) corresponds to an avoided crossing of quasi energies, which is observed at odd resonances ($N$ odd). In this case,  as $\epsilon$ is continuously varied to cross the resonance, the Floquet eigenstates in each branch are flipped, from level-1 to level-2 dominant (or viceversa).}
\end{figure}

\section{Nonadiabatic transitions}
The two-level equations (12) and (13) with periodic coefficients provide the starting point to demonstrate breakdown of adiabatic following for special driving frequencies $\epsilon$ when the dynamics is observed for time scales longer than the oscillation cycle $\sim 1 / \epsilon$. After setting $\mathbf{a}(t)=(a_1(t),a_2(t))^T$, Floquet theory states that the solution to Eqs.(12) and (13) with the initial condition $\mathbf{a}(0)=(1,0)^T$ is given by
\begin{equation}
\mathbf{a}(t)=\Phi(t) \exp(-i \mathcal{R}t) \mathbf{a}(0)
\end{equation}
  where $\Phi(t+2 \pi / \epsilon)=\Phi(t)$ is a $2 \times 2$ periodic matrix, with $\Phi(0)=1$, and the two eigenvalues $\mu_1$ and $\mu_2$ of the Floquet matrix $\mathcal{R}$ define the quasi energies (Floquet exponents).  The quasi energies are defined apart from integer multiplies than the oscillation frequency $\epsilon$. Therefore, quasi energy degeneracy is attained whenever the difference $\mu_2-\mu_1$ is an integer multiple than $ \epsilon$. A different way to write Eq.(15) is to introduce the Floquet eigenstates associated to the quasi energies $\mu_1$ and $\mu_1$. Indicating by $\mathbf{q}_1$ and $\mathbf{q}_2$ the eigenvectors of $\mathcal{R}$ corresponding to the eigenvalues $\mu_1$ and $\mu_2$, the Floquet eigenstates are defined by $\mathbf{W}_1(t)= \Phi(t) \mathbf{q}_1$ and  $\mathbf{W}_2(t)=\Phi(t) \mathbf{q}_2$. Provided that $\mathbf{q}_1$ and $\mathbf{q}_2$ are linearly independent, the solution $\mathbf{a}(t)$ can written as a superposition of Floquet states with coefficients $\alpha$ and $\beta$, namely
  \begin{equation}
  \mathbf{a}(t)=\alpha \mathbf{W}_1(t) \exp(-i \mu_1 t)+ \beta \mathbf{W}_2(t) \exp(-i \mu_2 t).
  \end{equation}

  The values of $\alpha$ and $\beta$ are determined such as to satisfy the initial condition $\mathbf{a}(0)=(1,0)^T$.
   Note that the Floquet eigenstates $\mathbf{W}_1(t)$ and $\mathbf{W}_2(t)$ are periodic functions of time with period $ 2 \pi / \epsilon$. Note also that, while Eq.(15) is always a valid result, Eq.(16) fails to describe the correct dynamics when the matrix $\mathcal{R}$ becomes defective, i.e. at an EP where both quasi energies and the Floquet eigenstates coalesce. This singular case can occur for non-Hermitian shaking of the potential well and will be discussed further in the following Sec. III.B.\\
   The quasi energies $\mu_1, \mu_2$ and corresponding Floquet states can be computed by standard methods; see Appendix A for technical details. For Hermitian shaking, the quasi energies are real, however for non-Hermitian shaking they can become complex. The appearance of complex quasi energies makes the dynamics rather trivial, since the Floquet state corresponding to the quasi energy with the largest imaginary part becomes the dominant mode after some time. Therefore, here we will limit to consider the case of non-Hermitian shaking but with real quasi energies. For the modulation function $f(\epsilon t)$ defined by Eq.(14), it turns out that the quasi energies are real provided that $V_1V_2$ is real (see Appendix A), and the quasi energies can be chosen to satisfy the condition $\mu_2=-\mu_1$. Moreover, for $V_1V_2 \neq 0$ the dynamics is pseudo-Hermitian, i.e. it can be reduced to an equivalent Hermitian dynamics with a sinusoidal shaking of the potential in real space. Therefore in the following we can limit ourselves to consider the two different cases (i) $V_1=V_2=A/2$ real (Hermitian shaking), and (ii) $V_1=0$.
      
\subsection{Hermitian shaking: nonadiabatic transitions and Rabi flopping at multiphoton resonances}
For a sinusoidal shaking of the potential well in real space, $x_0(\epsilon t)=A \sin (\epsilon t)$, one has $V_1=V_2=A \kappa/2$ and breakdown of adiabatic following for the driven two-level model [Eqs.(12) and (13)] is observed close to Floquet quasi-degeneracies \cite{r35}. In the spirit of the adiabatic limit $\epsilon \rightarrow 0$, an approximate expression of the quasi energies and corresponding Floquet eigenstates can be obtained by a standard WKB analysis  of Eqs.(12) and (13) (see, for instance, \cite{r35,r58,r59}). This yields
\begin{equation}
\mu_{1,2} \simeq \mp \frac{\epsilon}{2 \pi} \int_0^{2 \pi / \epsilon} dt \sqrt{\left( \frac{\omega_0}{2}\right)^2+\epsilon^2 f^2 (\epsilon t)}
 \end{equation}
   \begin{widetext}
    \begin{equation}
   \mathbf{W}_1(t) \simeq  \mathbf{W}_1^{(WKB)}(t) \equiv \frac{1}{\omega_0}\left( 
   \begin{array}{c}
    \omega_0/2+\lambda(\epsilon t) \\
   -\epsilon f(\epsilon t) 
   \end{array}
   \right) \exp \left[ i \int_0^t d \eta \lambda(\epsilon \eta) +i \mu_1 t \right]
   \end{equation}
   \begin{equation}
   \mathbf{W}_2(t) \simeq   \mathbf{W}_2^{(WKB)}(t) \equiv \frac{1}{\omega_0}\left( 
   \begin{array}{c}
   \epsilon f(\epsilon t) \\
   \omega_0/2+\lambda(\epsilon t) 
   \end{array}
   \right) \exp \left[ -i \int_0^t d \eta \lambda(\epsilon \eta) +i \mu_2 t \right]
   \end{equation}
     \end{widetext}
where we have set $\lambda(\epsilon t) \equiv \sqrt{(\omega_0/2)^2+\epsilon^2 f^2 (\epsilon t)}$. Note that, at leading order in $\epsilon$, apart from a phase factor one has 
$\mathbf{W}_1^{(WKB)}(t) \sim (1, O(\epsilon) )^T$, $\mathbf{W}_2^{(WKB)}(t) \sim ( O(\epsilon) ,1)^T$and 
\begin{equation}
\mu_{1,2}  \simeq \mp \frac{\omega_0}{2} \left( 1+\frac{2V_1V_2 \epsilon^2}{\omega_0^2} \right) +O(\epsilon^4),
\end{equation}
 i.e. within the limits of validity of the WKB approximation $\mathbf{W}_1(t)$ is level-1 dominant whereas $\mathbf{W}_2(t)$ is level-2 dominant. Note that $\mu_1$ ($\mu_2$) is a decreasing (increasing) function of $\epsilon$. If the quasi energies $\mu_1$ and $\mu_2$ are far from being degenerate, the WKB analysis provides an accurate estimate of the Floquet eigenstates and, according to Eq.(16), one should choose $\alpha \simeq 1$ and $\beta \simeq 0$. Therefore, far from quasi energy degeneracies adiabatic following is expected for an arbitrarily long time. Possible breakdown of adiabatic following can be observed close to quasi energy degeneracies, i.e. when the difference $(\mu_2-\mu_1)$ is an integer multiple than $\epsilon$ (see also the recent study \cite{r35}). The values of the oscillation frequency $\epsilon$ corresponding to (near) quasi energy degeneracy can be estimated from the WKB form of the Floquet exponents [Eq.(17)] by imposing
\begin{equation}
\int_0^{2 \pi / \epsilon} dt \sqrt{\left( \frac{\omega_0}{2} \right)^2+ \epsilon^2 f^2( \epsilon t) } =N \pi
\end{equation}
where $N$ is a (sufficiently large)  integer number. Substitution of Eq.(14) into Eq.(21), at leading order in $\epsilon$ one obtains the following values $\epsilon=\epsilon_N$ of oscillation frequencies for quasi energy degeneracy
\begin{equation}
\epsilon_N \simeq \frac{\omega_0}{N} \left( 1+\frac{4 V_1 V_2}{N^2} \right).
\end{equation}
When the oscillation frequency $\epsilon$ is close to $\epsilon_N$, the exact form of the Floquet eigestates can not be predicted by the WKB analysis, since near the degeneracy point a mixing of $\mathbf{W}_1^{(WKB)}(t)$ and $\mathbf{W}_2^{(WKB)}(t)$  is  possible, and we do not know a priori what is the right linear combination of $\mathbf{W}_1^{(WKB)}(t)$ and $\mathbf{W}_2^{(WKB)}(t)$ that gives the exact form of Floquet eigenstates. However, some general considerations can be drawn by considering the behavior of the quasi energies $\mu_2=-\mu_1$ versus $\epsilon$ near $\epsilon_N$. Two cases can be found, which are summarized in Fig.2. In the former case, which is observed at even resonances [Eq.(22) with $N$ even],  there is a crossing of quasi energies, corresponding to exact degeneracy of the quasi energies at $\epsilon= \epsilon_N$ (Hermitian degeneracy). Since the behavior of Floquet eigenstates varies continuously with $\epsilon$ near $\epsilon_N$ and since far from $\epsilon=\epsilon_N$ the two Floquet eigenstates are level-1 and level-2 dominant states [according to Eqs.(18) and (19)], there is not any mixing of states (18) and (19), and adiabatic following is again expected in this case [Fig.2(a)]. The other case corresponds to an avoided crossing of quasi energies [Fig.2(b)], which is observed at odd resonances $\epsilon_N$ [Eq.(22) with $N$ odd]. In this case a mixing of states (18) and (19) near $\epsilon=\epsilon_N$  is necessary to ensure continuous change of the Floquet eigenstates, from dominant level-1 to dominant level-2, in each quasi energy branch, as schematically shown in Fig.2(b). The exact Floquet eigenstates near $\epsilon \simeq \epsilon_N$ are thus given by linear combinations $\mathbf{W}_1(t)= \gamma_{11} \mathbf{W}_1^{(WKB)}(t)+\gamma_{12} \mathbf{W}_2^{(WKB)}(t)$ and
$\mathbf{W}_2(t)= \gamma_{21} \mathbf{W}_1^{(WKB)}(t)+\gamma_{22} \mathbf{W}_2^{(WKB)}(t)$ of WKB eigenstates with suitable $\gamma_{i,j}$ coefficients, which  rapidly change as the avoided crossing point is swept. As a result, the exact Floquet eigenstates near the quasi-degeneracy point are neither level-1 nor level-2 dominated. In particular, for $\gamma_{11}=\gamma_{12}=\gamma_{21}=1/ \sqrt{2}$, $\gamma_{22}=-1/ \sqrt{2}$ fully mixing of level occupation is obtained, and in Eq.(16) one has to assume $\alpha \simeq \beta \simeq 1/ \sqrt{2}$ to satisfy the initial condition. 
The evolution of amplitudes $a_1(t)$ and $a_2(t)$ is governed by the interference of the two Floquet eigenstates with phase mismatch $(\mu_2-\mu_1)t$.  Owing to the non vanishing separation $2 \Delta$ of quasi energies  at the avoided level crossing [Fig.2(b)], the phase mismatch leads to alternating in-phase and out-of-phase superposition of the exact Floquet eigenstates, corresponding to Rabi flopping between levels 1 and 2 at the Rabi frequency $\Omega_{R}= \pi / \Delta$.\\ 
Figure 3(a) shows, as an example, the numerically-computed behavior of the quasi energies versus normalized oscillation frequency $\epsilon / \omega_0$ for  $V_1=V_2=0.5$, clearly showing Hermitian degeneracy and avoided crossing at even and odd resonances, respectively. The rapid change of Floquet eigenstates, from dominant level-1 to dominant level-2, near the odd resonances (avoided crossing) is shown in Fig.3(b), which depicts the behavior of the unbalance factor $\theta$ versus $\epsilon / \omega_0$. The unbalance factor is defined as 
\begin{equation}
\theta=\left| {\rm max}|A_n|-{\rm max}|B_n| \right|,
\end{equation}
\begin{figure}[htbp]
  \includegraphics[width=87mm]{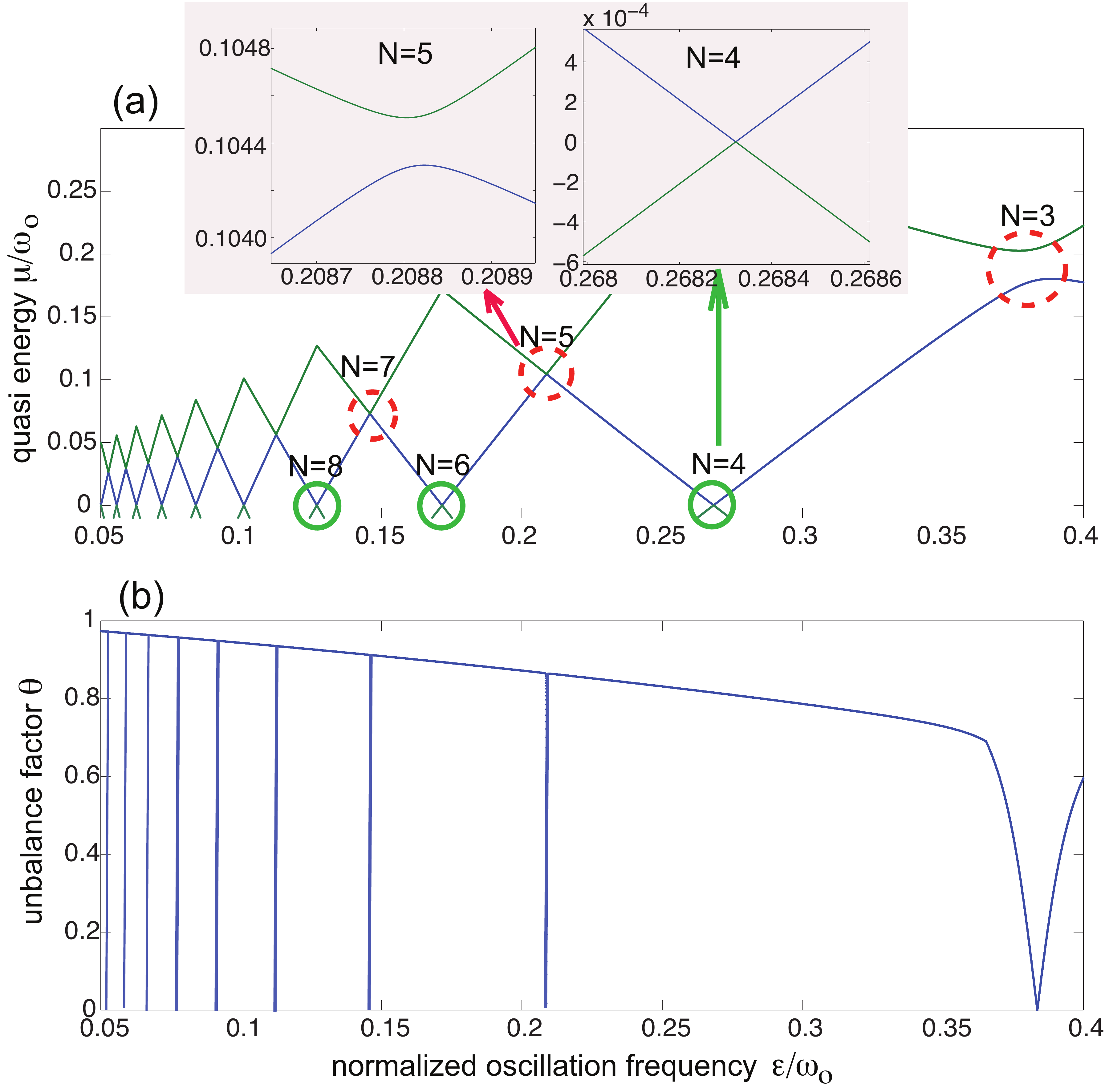}\\
   \caption{(color online) (a) Numerically-computed behavior of quasi energies $\mu_{1,2}$ versus oscillation frequency $\epsilon$ for the reduced  two-level model [Eqs.(12) and (13)] and for a sinusoidal Hermitian shaking $f(\epsilon t)=2 V \cos (\epsilon t)$ with amplitude $V=0.5$. Quasi energies are plotted in the range $(0^-, \epsilon)$. Even resonances (solid circles) correspond to quasi energy level crossing (Hermitian degeneracies), whereas odd resonances (dashed circles) correspond to avoided crossing of quasi energies. Detailed behavior of quasi energies near the $N=4$ and $N=5$ resonances is shown in the upper insets. (b) Behavior of the unbalance factor $\theta$ versus oscillation frequency. The abrupt and narrow drops of the unbalance factor toward zero at the odd resonances is the signature of rapid change of Floquet eigenstates from dominant level-1/level-2 states ($\theta$ large and close to one) to balanced level occupation ($\theta$ close to zero).}
\end{figure}
 where $A_n$ and $B_n$ are the Fourier components of either one of the Floquet eigenstates $\mathbf{W}_1(t)$ or $\mathbf{W}_2(t)$ (see Appendix A). A value of $\theta$ close to one means that the Floquet eigenstates are level-1 and level-2 dominant, according to the WKB analysis. On the other  hand, a value of $\theta$ close to zero means that the occupation of the two levels in the Floquet eigenstates is balanced. An inspection of Fig.3(b) clearly shows that, far from the odd resonances $N=3,5,7,...$, $\theta$ is almost close to one (at least for large $N$), indicating that the Floquet eigenstates are either level-1 or level-2 dominant and well approximated by $\mathbf{W}_{1,2}^{(WKB)}(t)$. Conversely, close to the odd resonances abrupt and very narrow  drops of $\theta$ to zero are observed, indicating that at avoided quasi energy crossing the Floquet eigenstates equally populate the two levels. Figure 4 shows typical examples of the two-level dynamics in time domain for an oscillation frequency that spans either an odd resonance [$N=5$, Fig.4(a)] or an even resonance [Fig.4(b), $N=4$]. The results are obtained by numerical simulations of the two-level equations (12) and (13) using an accurate fourth-order Runge-Kutta method with variable step, for a modulation function $f(t)=2 V \cos (\epsilon t)$ with $V=0.5$ and with the initial condition $a_1(0)=1$, $a_2(0)=0$. Note that in the latter case [Fig.4(b), even resonance] the system remains almost in the initial level for extremely long times, i.e. adiabatic following is observed well beyond the time scale $1 / \epsilon$. Conversely, for an odd resonance [Fig.4(a)] nonadiabatic transitions are clearly observed in the form of Rabi oscillations between levels 1 and 2 when the oscillation frequency $\epsilon$ crosses the resonance frequency $\epsilon_N$. The frequency of Rabi oscillations observed in the numerical simulations turns out to be in excellent agreement with the theoretical value $\Omega_R= \pi / \Delta$ predicted by the Floquet analysis, where $ 2 \Delta$ is the separation of the quasi energies at the avoided level crossing. From a physical viewpoint, breakdown of the QAT for the Hermitian shaking of the potential well can be explained as a result of a multiphoton absorption process, from the ground state $E_1$ to the excited state $E_2$, yielding multiphoton Rabi oscillations between the two levels \cite{r36,r37,r38}. A similar phenomenon has been predicted for a periodically-driven two-level model in Ref.\cite{r35} and suggested to be a rather universal phenomenon of periodically and slowly-changing Hermitian Hamiltonians.
\subsection{Non-Hermitian shaking: nonadiabatic transitions near Floquet exceptional points}  
Let us consider a non-Hermitian shaking of the potential well corresponding to $V_1=0$. Such a case is obtained by assuming the oscillation path $x_0(\epsilon t)=A-A\exp(-i \epsilon t)$, i.e. $A_1=0$ and $A_2=i A$ in Eq.(7), yielding $V_1=0$ and $V_2=i \kappa A$. In this case it can be shown (see Appendix B) that the {\it exact} quasi energies are given by
\begin{equation}
\mu_1=-\frac{\omega_0}{2} \; ,\;\; \mu_2=\frac{\omega_0}{2}
\end{equation}
so that exact quasi energy level crossing ($\mu_2-\mu_1=N \epsilon$) is found at the resonance frequencies $\epsilon=\epsilon_N$ with
\begin{equation}
\epsilon_N = \frac{\omega_0}{N}.
\end{equation}

  \begin{figure*}[htbp]
  \includegraphics[width=170mm]{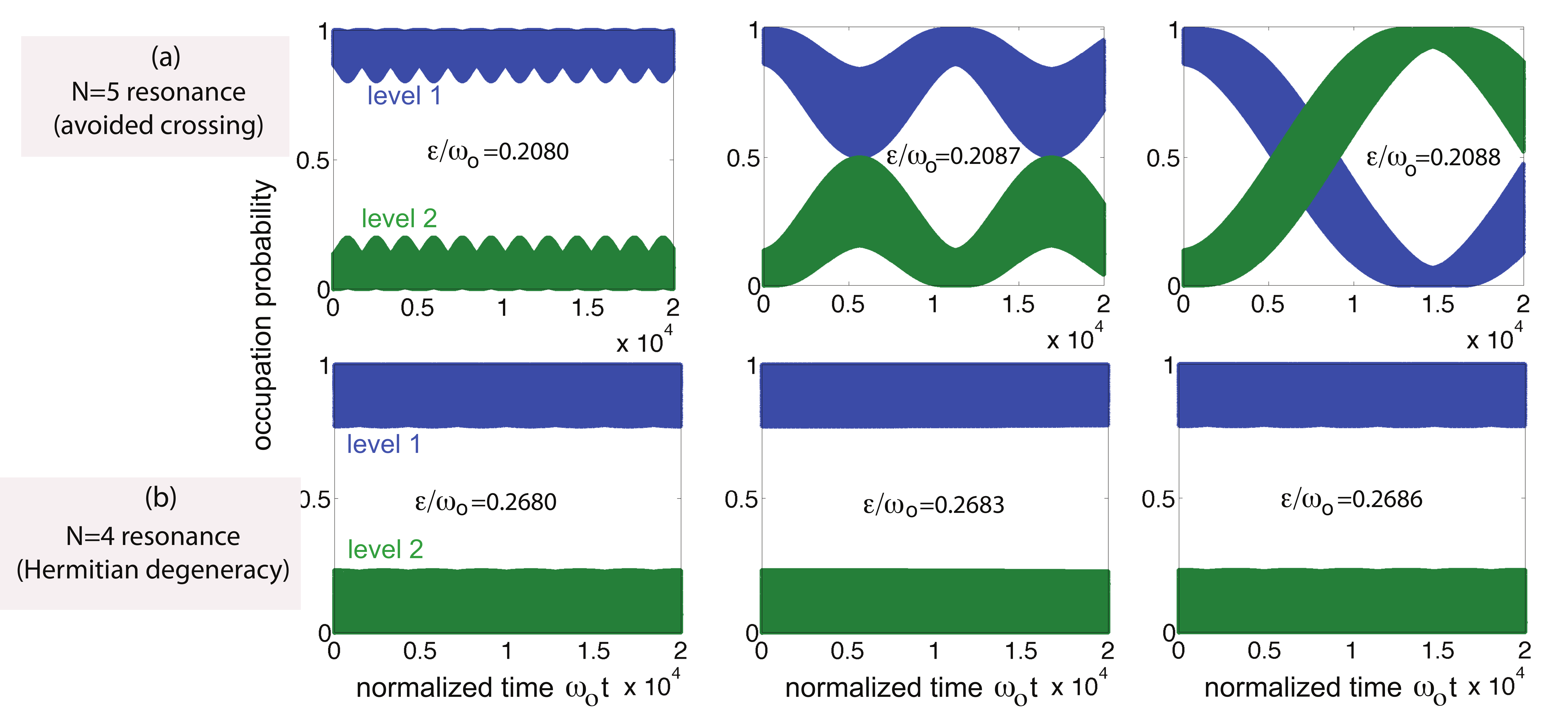}\\
   \caption{(color online) Numerically-computed evolution of the level occupation probabilities $|a_1(t)|^2$ and $|a_2(t)|^2$ for the reduced  two-level model [Eqs.(12) and (13)] and for a sinusoidal shaking $f(\epsilon t)=2 V \cos (\epsilon t)$ with amplitude $V=0.5$. At initial time the particle occupies level 1. Panels in (a) show the dynamical behavior when the oscillation frequency $\epsilon$ spans the $N=5$ resonance, corresponding to avoided crossing of quasi energies, whereas panels in (b) show the dynamical behavior when $\epsilon$ spans the $N=4$ resonance (Hermitian degeneracy). In the former case failure of the QAT is clearly observed owing to the appearance of Rabi oscillations. In the left and central panels of (a) the Rabi oscillations are detuned, whereas in the right panel they are at resonance. The frequency of Rabi oscillations at resonance  turns out to be $\Omega_R \simeq 2.14 \times 10^{-4} \omega_0$, which is in agreement with the result $\Omega_R= \pi / \Delta$ predicted by Floquet theory, where $2 \Delta$ is the separation of the quasi energies at the avoided level crossing for the $N=5$ resonance [see the inset in Fg.3(a)].}
\end{figure*}

  \begin{figure*}[htbp]
  \includegraphics[width=180mm]{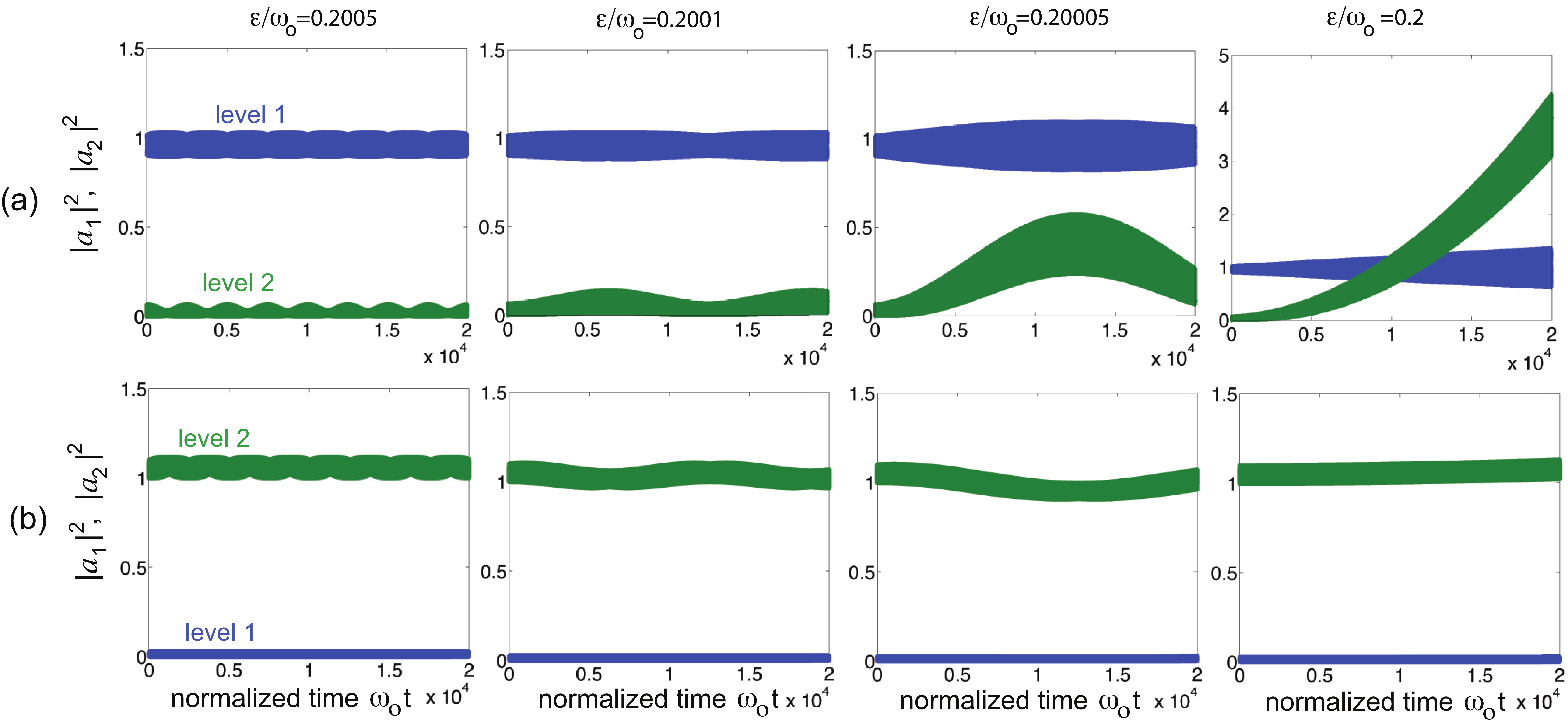}\\
   \caption{(color online) (a) Breakdown of the QAT near an EP for the two-level model Eqs.(12) and (13) and for the non-Hermitian shaking $f(\epsilon t)= iV \exp (-i\epsilon t)$, with amplitude $V=0.5$. The four panels show the numerically-computed evolution of $|a_1(t)|^2$ and $|a_2(t)|^2$  for a set of oscillation frequencies close to the $n=5$ odd resonance, clearly showing breakdown of adiabatic following as the EP $\epsilon=\omega_0/n$ is approached. The initial condition is $a_1(0)=1$ and $a_2(0)=0$. (b) Same as (a), but for the initial condition $a_1(0)=0$ and $a_2(0)=1$.}
\end{figure*}

For $\epsilon$ far from any odd resonance, the Floquet eigenstates are linear independent, with $\mathbf{W}_1(t)$ and $\mathbf{W}_2(t)$ being level-1 and level-2 dominant, respectively. However, as $\epsilon$ approaches an odd resonance, i.e. $\epsilon \rightarrow \omega_0/(2N-1)$, the coalescence of quasi energies is associated to a simultaneous coalescence of Floquet eigenstate, with $\mathbf{W}_1(t)$ showing a rather abrupt change and becoming level-2 dominant like $\mathbf{W}_2$ (see  Appendix B). This means that, at an odd resonance $\epsilon=\epsilon_{2N-1}$, the non-Hermitian periodic two-level Hamiltonian, defined by Eqs.(12) and (13), shows a Floquet EP. Here we wish to show how the appearance of an EP breaks the QAT. To this aim, let us first notice that the evolution of the amplitudes $a_1(t)$ and $a_2(t)$, as described by Eq.(15), can be mapped into an equivalent evolution of a non-Hermitian time-independent Hamiltonian at discretized times $t=0,T_{\epsilon},2T_{\epsilon},3T_{\epsilon},...$, where $T_{\epsilon}= 2 \pi / \epsilon$ is the oscillation period. In fact, at $t=nT_{\epsilon}$ ($n=0,1,2,3,...$) from Eq.(15) one has
\begin{equation}
\mathbf{a}(nT_{\epsilon})=\exp(-i \mathcal{R}nT_{\epsilon}) \mathbf{a}(0)
\end{equation}
where we used the property $\Phi(n T_{\epsilon})=1$. The exponential matrix $\exp(-i \mathcal{R}nT_{\epsilon})$ on the right hand side of Eq.(26) can be viewed as the propagator, over a time $nT_{\epsilon}$, of the time-independent Hamiltonian $\mathcal{R}$, i.e. Eq.(26) can be viewed as the solution to the Schr\"odinger equation
\begin{equation}
i \frac{d \mathbf{a}}{dt}= \mathcal{R} \mathbf{a}(t).
\end{equation}
At an EP, the Floquet matrix $\mathcal{R}$ is defective, i.e. the two eigenvalues of $\mathcal{R}$ (quasi energies) and corresponding eigenvectors coalesce \cite{r49}. As discussed in Appendix B, the (unique) eigenvector $\mathbf{q_2}$, satisfying the equation $\mathcal{R}\mathbf{q_2}= (\omega_0/2) \mathbf{q_2}$, corresponds to a level-2 dominant Floquet eigenstate, i.e. $\mathbf{q_2} \sim (0,1)^T$. An associated (or generalized) eigenvector $\mathbf{Q}_2$ of the defective matrix $\mathcal{R}$ can be then introduced \cite{r49} by solving the matrix equation
\begin{equation}
\left( \mathcal{R}-\frac{\omega_0}{2} \right) \mathbf{Q}_2=\mathbf{q}_2.
\end{equation}
The associated eigenvector $\mathbf{Q}_2$ corresponds to level-1 dominant state, i.e. $\mathbf{Q}_2 \sim (1,0)^T$.
The most general solution to Eq.(27) is given by
\begin{equation}
\mathbf{a}(t)=\left[ (-i \gamma t +\delta)\mathbf{q}_2+ \gamma \mathbf{Q}_2 \right] \exp(-i \omega_0 t /2)
\end{equation}
 as one can readily check by direct calculations.
The constants $\gamma$ and $\delta$ are determined by the initial condition $\mathbf{a}(0)=(1,0)^T \sim \mathbf{Q}_2$, i.e. $\gamma \sim 1$ and $\delta \sim 0$. Equation (29) clearly shows 
that, at long times, $\mathbf{a}(t)$ is dominated by the secularly-growing term $ \sim -i t \mathbf{q}_2$, and thus level $E_2$ becomes more occupied than level $E_1$ indicating breakdown of the QAT. Figure 5(a) shows, as an example, breakdown of the QAT for non-Hermitian shaking near an EP as obtained by numerical simulations of the two-level equations (12) and (13) for a modulation function $f(\epsilon t)=iV \exp(-i \epsilon t)$, corresponding to the non-Hermitian shaking $x_0(\epsilon t)=A-A \exp(-i \epsilon t)$ of the potential well with $V=\kappa A$. The above analysis also indicates that, if the system is initially prepared in level 2 (rather than in level 1), i.e. for $a_1(0)=0$ and $a_2(0)=1$, level 2 remains the dominant one in the dynamics and nonadiabatic transitions are prevented: in fact, with such an initial condition one should take in $\gamma \sim 0$ and $\delta \sim 1$ in Eq.(29), so that  $|a_1(t) / a_2(t)|^2$ remains small for growing time $t$. This behavior is confirmed by direct numerical simulations, as shown in Fig.5(b). Such an asymmetric behavior, i.e. the appearance of nonadiabatic transitions for the system initially prepared in one of the two levels, but not in the other one, is a peculiar feature of non-Hermitian dynamics without any counterpart in Hermitian systems \cite{r42,r43,r47,r48}. From a physical viewpoint, asymmetric breakdown of the QAT can be  explained by observing that  the modulation (coupling) function $f(\epsilon t)=iV \exp(-i \epsilon t)$ has a one-sided Fourier spectrum, i.e. it is composed by negative-frequency components solely, so that it can induce only `upward` transitions, i.e. transitions to higher energy levels \cite{r60,r61}. Therefore, while a multiphoton transition from the ground level $E_1$ to the excited level $E_2$ is allowed, transition from level $E_2$ to level $E_1$ is forbidden for non-Hermitian shaking.

\section{An example: periodically-shaken double-well potential}
As an illustrative example, we consider a periodically-shaken double-well potential sustaining two bound states solely with energies $E_1=-\sigma_1^2$ (ground state) and  $E_2=-\sigma_2^2$ (excited state), spaced by the energy $\omega_0=\sigma_1^2-\sigma_2^2$; see Fig.1(b). The potential well can be synthesized by supersymmetric quantum mechanics and reads explicitly \cite{r62}
\begin{equation}
V(x)=\frac{-2 \omega_0 \left[ \sigma_1^2 \cosh^2 (\sigma_2 x)+\sigma_2^2 \sinh^2( \sigma_1 x)  \right]}{\left[ \sigma_2 \sinh (\sigma_1 x) \sinh(\sigma_2 x) - \sigma_1 \cosh(\sigma_1 x) \cosh (\sigma_2 x) \right]^2}.
\end{equation}
The eigenfunctions $u_1(x)$ and $u_2(x)$, corresponding to the energies $E_1=-\sigma_1^2$ and $E_2=-\sigma_2^2$, are given by
\begin{eqnarray}
u_1(x) & = & \frac{\mathcal{N}_1}{- \sigma_1 \cosh (\sigma_1 x)+ \sigma_2 \tanh (\sigma_2 x) \sinh (\sigma_1 x) } \\
u_2(x) & = & 	\frac{\mathcal{N}_2 \sinh (\sigma_1 x) }{\sigma_2 \sinh (\sigma_1 x) \sinh( \sigma_2 x)-\sigma_1 \cosh (\sigma_1 x) \cosh (\sigma_2 x) } \;\;\;\;
\end{eqnarray}
\begin{figure}[htbp]
  \includegraphics[width=85mm]{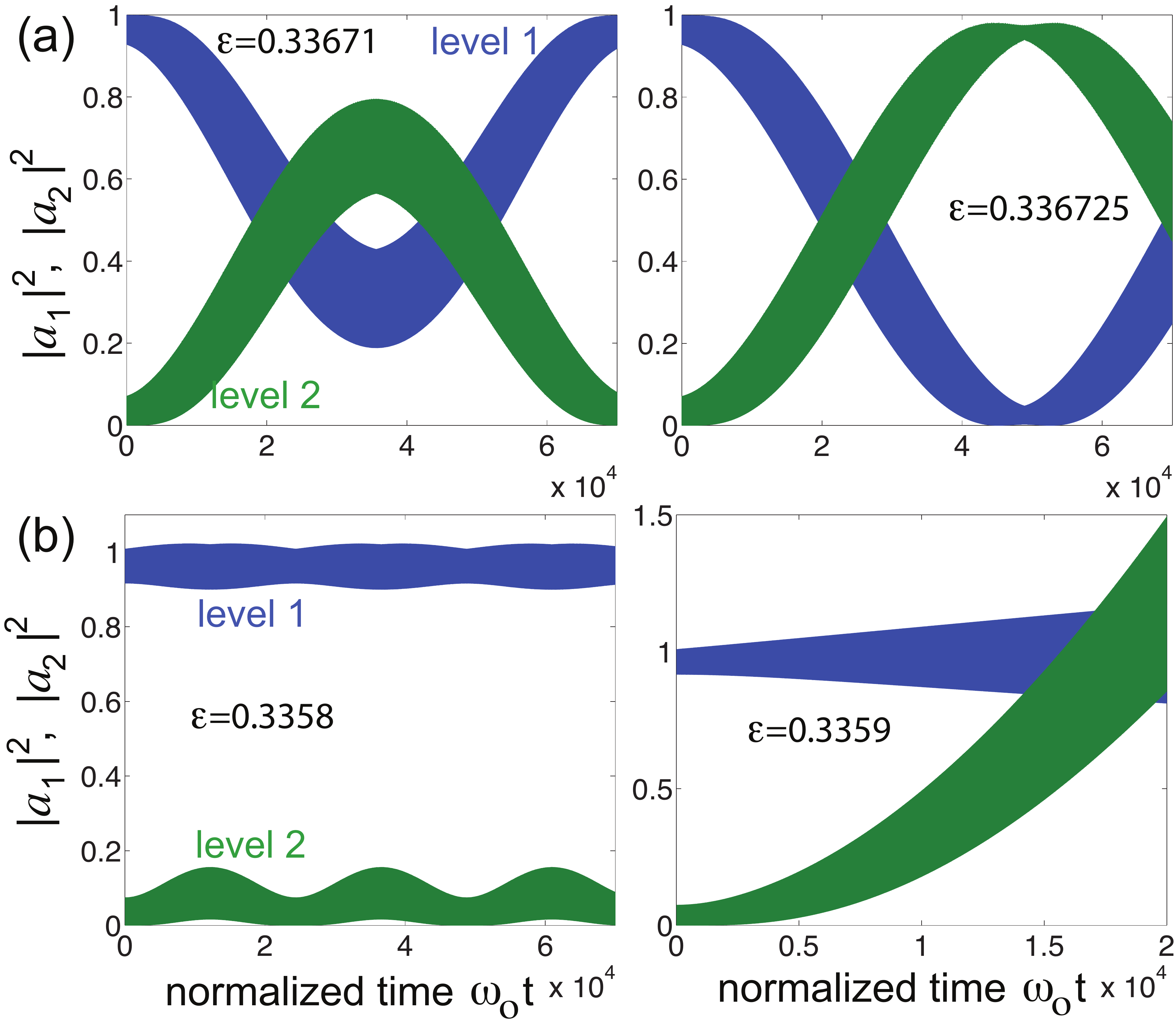}\\
   \caption{(color online) Behavior of $|a_{1}(t)|^2$ and $|a_2(t)|^2$ for the periodically-shaken double-well potential of Fig.1(b) as obtained by numerical integration of the Schr\"odinger equation (1) using a pseudo-spectral split-step method. Integration domain: $ -8<x<8$, space discretization $dx=0.0627$, time discretization $dt=0.01$. (a) Sinusoidal Hermitian shaking $x_0(\epsilon t)= \sin (\epsilon t)$, (b) non-Hermitian shaking $f(\epsilon t)=0.6-0.6 \exp(-i \epsilon t)$. Oscillation frequencies are set close to the three-photon resonance $\epsilon \simeq \omega_0/3 =1/3$.}
\end{figure}
 where $\mathcal{N}_1$ and $\mathcal{N}_2$ are normalization constants. We checked breakdown of adiabaticity for either Hermitian and non-Hermitian periodic shaking of the double-well potential by numerical integration of the Schr\"odinger equation (1) using a standard pseudo-spectral split-step method for parameter values $\sigma_2=\sqrt{2}$ and $\sigma_1=\sqrt{3}$. The oscillation frequency $\epsilon$ was chosen close to the third-order resonance $ \epsilon \simeq \omega_0 /3=1/3$. At initial time the wave function $\psi(x,0)$ is set equal to the ground level eigenfunction $u_1(x)$, and the evolution of the occupation level amplitudes 
 \begin{eqnarray}
 a_1(t) = \int dx u_1(x-x_0(\epsilon t)) \psi(x,t) dx \; , \nonumber \\
 a_2(t) = \int dx u_2(x-x_0(\epsilon t)) \psi(x,t) dx
 \end{eqnarray}
  are computed up to the long time scale $t \sim 7 \times 10^4$.  Figure 6(a) shows the numerical results corresponding to the Hermitian shaking $x_0(\epsilon t)=A \sin (\epsilon t)$ with $A=1$, clearly showing breakdown of the QAT owing to multiphoton Rabi oscillations. Note that during the dynamics most of the excitation remains  in either level 1 or level 2, excitation to the continuum of states (ionization) being negligible. Such a result justifies the approximation made in Sec.II to neglect the continuum of states and the consider a two-level model. Breakdown of the QAT for the non-Hermitian shaking  $x_0(\epsilon t)=A-A \exp (-i\epsilon t)$ is shown in Fig.6(b) for an oscillation amplitude $A=0.6$. Note that a secular growth of amplitude $a_2(t)$  is observed at the oscillation frequency $\epsilon=0.3359$, which is the signature of the Floquet EP. This value of $\epsilon$ turns out to be slightly detuned from the one predicted by the theoretical analysis $\epsilon=1/3$, probably due to a slight deviation of the energy level separation of the potential well from the theoretical one $\omega_0=1$ arising from space-time discretization of the Schr\"odinger equation in the numerical analysis.  
  
  \section{Application of non-Hermitian shaking to perturbative mode selection}
 As a simple physical application of non-Hermitian shaking and breakdown of the adiabatic theorem arising from a Floquet EP, we briefly discuss light mode selection in an optical directional coupler, made of two evanescently-coupled straight optical waveguides, induced by a {\em perturbative} periodic longitudinal modulation of complex refractive index. Coupled optical waveguide structures, including the optical directional coupler system, have been often used to emulate in photonics a wealth of quantum phenomena in the matter \cite{r55,refe1,refe2,r63}. Here we focus to mode selection in a directional coupler \cite{r55,r63}, however our simple model could be applied to realize mode selection in other effective two-level systems, such as in two coupled optical microrings with temporal modulation of their complex resonance frequencies.\\
 Indicating by $b_1(z)$ and $b_2(z)$ the amplitudes of light waves trapped in the two waveguide modes, evolution of the light field along the longitudinal propagation direction $z$ of the coupler is governed by coupled-mode equations \cite{r55,refe2,r63}
 \begin{eqnarray}
 i \frac{db_1}{dz}= -\kappa_e b_2+\epsilon f(z) b_1 \\
 i \frac{db_2}{dz}= -\kappa_e b_1-\epsilon f(z) b_2  
 \end{eqnarray} 
where $\kappa_e$ is the coupling constant between waveguide modes due to evanescent coupling and $\epsilon f(\epsilon z)$ describes a small and slowly-varying change, along the longitudinal propagation distance $z$, of the effective mode index (real and imaginary parts) in the two waveguides. Here $\epsilon$ is a dimensionless parameter that measures the smallness of the change of the effective mode index as compared to the unperturbed one in the waveguides. The real part of $\epsilon f(\epsilon z)$ describes a change of the effective propagation constant arising from a modulation of the real part of the refractive index, whereas the imaginary part of $\epsilon f(\epsilon z)$ accounts for amplification or attenuation of the optical field due to optical gain or loss. Note that the modulation of the effective mode index is assumed antisymmetric in the two waveguides. Antisymmetric and periodic variation of the real part of the effective refractive index in the coupler, i.e. of the real part of $\epsilon f (\epsilon z)$, can be obtained by suitable periodic bending of the waveguide axis \cite{r55,r63bis}, whereas optical loss and gain controlling the imaginary part of $\epsilon f(\epsilon z)$ can be provide by selective optical absorption and optical gain in the structure.\\ 
To study the evolution of the light beam in the directional coupler, it is worth 
projecting the dynamics into the symmetric (S) and antisymmetric (A) supermodes of the coupler via the transformation
\begin{equation}
a_1(z)=\frac{b_1(z)+b_2(z)}{\sqrt{2}} \; ,\;\;  a_2(z)=\frac{b_1(z)-b_2(z)}{\sqrt{2}}.
\end{equation}
The amplitudes $a_1$ and $a_2$ of S and A supermodes thus satisfy the coupled-mode equations
 \begin{eqnarray}
 i \frac{da_1}{dz}= -\kappa_e a_1+\epsilon f(\epsilon z) a_2 \\
 i \frac{da_2}{dz}= \kappa_e a_2+\epsilon f(\epsilon z) a_1.  
 \end{eqnarray} 
Note that, after the formal substitution $z \rightarrow t$ and $\kappa_e \rightarrow \omega_0/2$, Eqs.(37) and (38) and formally equivalent to the two-level equations (12) and (13) of the periodically-shaken quantum potential in complex plane. Therefore, assuming a non-Hermitian modulation of the form (14) with $V_1=0$, according to the results of Fig.5 at a Floquet EP the dominant mode is the A mode (level 2). This means that, regardless of the initial light excitation of the coupler, the small (perturbative) modulation of the complex refractive index along the propagation direction enforces the antisymmetric mode. Non-Hermitian shaking at a Floquet EP thus provides a means to realize perturbative mode selection in the coupler.

  \section{Conclusions}
 The quantum adiabatic theorem is a cornerstone in quantum physics, which finds important applications in different areas of quantum physics. In its simplest version, it states that a quantum system, initially prepared in the ground state, evolves remaining  in the instantaneous ground state when the Hamiltonian is slowly changed in time [$\hat{H}=\hat{H}(\epsilon t)$ with $\epsilon \rightarrow 0$], provided that the instantaneous ground energy level remains separated from the other energy levels by a finite gap. However, such a prediction holds when the system is observed up to a log time scale of order $ \sim 1/ \epsilon$. At longer time scales,  nonadiabatic transitions can be observed, especially when the Hamiltonian contains oscillating terms \cite{r26,r35}. Breakdown of adiabatic evolution is even more striking when the Hamiltonian is described by a non-Hermitian operator \cite{r42,r43,r47,r48}, which can be experimentally realized in electromagnetic, electronic and optical systems \cite{r48,r63,r64,r65,r66,r67,r68,r69,r70,r71}. In this work we have shown that breakdown of adiabatic evolution can arise in a renowned model of quantum physics, namely in a periodically-shaken double-well potential \cite{r52}.  In ordinary Hermitian model, periodic shaking occurs in real space and can be exploited to either suppress or enhance quantum tunneling \cite{r52}. Here we extended the oscillation of the potential well into the complex plane, i.e. we considered a time-dependent potential $V(x,t)=V(x-x_0(t))$ with a spatial displacement $x_0(t)$ in either real space (Hermitian shaking, $x_0$ real) or in complex space (non-Hermitian shaking, $x_0$ complex). We have shown that for both Hermitian and non-Hermitian shaking of the potential well breakdown of the QAT is observed for long observation times whenever the oscillation frequency $\epsilon$ is tuned close to an odd resonance. However, the physical mechanism underlying nonadiabatic transitions is very distinct in the two cases. For the Hermitian shaking, nonadiabatic transitions arise from a multiphoton resonance process near avoided crossings of quasi energies and lead to Rabi flopping between the two levels,  with a mechanics similar to the one recently investigated in Ref.\cite{r35}. On the other hand, for the complex oscillating potential breakdown of the adiabatic theorem is rooted into the appearance of a Floquet EP, i.e. a singular regime where coalescence of both quasi energies and Floquet eigenstates occurs. Our results shed important physical insights into the long-time behavior of oscillating Hamiltonians. In particular, they show how breakdown of adiabatic evolution in non-Hermitian oscillating Hamiltonians can arise from the appearance of Floquet exceptional points, i.e. from the coalescence of both quasi energies and Floquet eigenstates, rather than from most common avoided crossing of quasi energies like in Hermitian oscillating Hamiltonians. 

  \appendix
\section{General properties of quasi energies and Floquet eigenstates}
The Floquet eigenstates and corresponding quasi energies can be found by looking for a solution to Eqs.(12) and (13) of the form
\begin{equation}
\left(
\begin{array}{c}
a_1(t) \\
a_2(t)
\end{array}
\right)=
\exp(-i \mu t) \sum_{n=-\infty}^{\infty}
\left(
\begin{array}{c}
A_n \\
B_n
\end{array}
\right) \exp(-i n \epsilon t)
\end{equation}
with $A_n,B_n \rightarrow 0$ as $|n| \rightarrow \infty$. Substitution of Ansatz (A1) into Eqs.(12) and (13) and using Eq.(14) yields the following hierarchical equations for the Fourier coefficients $A_n$ and $B_n$
\begin{eqnarray}
\left( \mu+n \epsilon+\frac{\omega_0}{2} \right)A_n= \epsilon (V_1 B_{n+1}+V_2 B_{n-1}) \\
\left( \mu+n \epsilon-\frac{\omega_0}{2} \right)B_n= \epsilon (V_1 A_{n+1}+V_2 A_{n-1}) 
\end{eqnarray}
The quasi energies $\mu$ can be viewed as the eigenvalues of an infinitely extended matrix. In practice, one truncates the index $|n|$ up to some large enough value $N$, i.e. one assumes $n=-N,...,N$ with $A_n=B_n \simeq 0$ for $|n|>N$, and calculate numerically $\mu$ as an eigenvalue of a $(2N \times 2 N)$ matrix. Since $\mu$ is defined apart from integer multiplies than $\epsilon$ and given the form of the hierarchical equations (A2) and (A3), there are no more than two distinct values of quasi energies, as it should. Let us now prove two properties of the quasi energies and Floquet eigenstates.\\
\\
{\textbf 1.} {\it  $A_n$, $B_n$ decay as $|n| \rightarrow \infty$ faster than exponential.}\\
Such a property readily follows by considering the asymptotic behavior of Eqs.(A1) and (A2) for large $|n|$, which yields the following recurrence relation for $A_n$
\begin{equation}
(V_1^2 A_{n+1}+V_2^2 A_{n-2})/n^2 \simeq  A_n 
\end{equation}
and a similar one for $B_n$. Such a recurrence relation shows that $|A_n|$ decays toward zero like $\sim (1/ n!)^2$ as $|n| \rightarrow \infty$. The same holds for $B_n$.\\
\\
{\textbf 2.} {\it If $V_1V_2$ is a real and non-vanishing number, then the quasi energies $\mu_1$ and $\mu_2$ are real and can be chosen to satisfy the condition $\mu_2=-\mu_1$}.\\
In fact, if $V_1V_2$ is a real and non vanishing number, we can set $V_1=|V_1| \exp(i \phi)$, $V_2=|V_2| \exp(-i \phi)$, with $\phi$ real and $|V_{1,2}| >0$. Let us make the substitution
\begin{equation}
A_n=\alpha_n \exp(i \theta n) \; , \; \; B_n=\beta_n \exp(i \theta n)
\end{equation}
where the complex angle $\theta$ is defined by the relation
\begin{equation}
\exp(i \theta) = \sqrt{\frac{|V_2|}{|V_1|}} \exp(-i \phi).
\end{equation}
Note that, since $A_n,B_n$ decay faster than an exponential as $|n| \rightarrow \infty$, the same decay behavior holds for the amplitudes $\alpha_n, \beta_n$, even though the imaginary part of $\theta$ is non vanishing.  After substitution of Eq.(A5) into Eqs.(A2) and (A3), one obtains
\begin{eqnarray}
\left( \mu+n \epsilon+\frac{\omega_0}{2} \right) \alpha_n= \epsilon \Gamma (\beta_{n+1}+\beta_{n-1}) \\
\left( \mu+n \epsilon-\frac{\omega_0}{2} \right) \beta_n= \epsilon \Gamma (\alpha_{n+1}+\alpha_{n-1}) 
\end{eqnarray}
where we have set $\Gamma \equiv \sqrt{|V_1 V_2|}$. In their present form, Eqs.(A7) and (A8) can be viewed as the hierarchical equations associated to the two-level equations (12) and (13) with the sinusoidal modulation function $f(\epsilon t)=2 \Gamma \cos (\epsilon t)$. Therefore, since the problem is Hermitian one, the quasi energies $\mu_1$ and $\mu_2$ should be real. Moreover, since $f(-\epsilon t)=f(\epsilon t)$, it follows that $\mu_2=-\mu_1$. In fact, if $\mathbf{W}_1(t)=(u(t), v(t))^T$ is a Floquet eigenstate with quasi energy $\mu_1$ for the modulation function $f(\epsilon t)=2 \Gamma \cos (\epsilon t)$, then it readily follows that $\mathbf{W}_2(t)=(v(-t),-u(-t))^T$ is a Floquet eigenstate as well with quasi energy $\mu_2=-\mu_1$.\\
The property 2. stated above shows that, for a non-Hermitian shaking of the potential well with $V_1V_2$ real and non vanishing number, the quasi energy spectrum is real despite the non-Hermitian  nature of the shaking (the potential $V(x,t)=V(x-x_0(\epsilon t))$ is complex). In this case the problem can be mapped {\it mutatis mutandis} to the Hermitian problem of the oscillating potential well in real space with a sinusoidal spatial displacement of appropriate amplitude. The non-Hermitian nature of the problem is accounted for by the renormalization of the Fourier amplitudes of the Floquet eigenstates according to Eq.(A5).

\section{Non-Hermitian shaking and Floquet exceptional points}
Let us consider a non-Hermitian shaking with $V_1=0$ and $V_2 \neq 0$, however a similar analysis could be done by taking $V_1 \neq  0$ and $V_2=0$. For $V_1=0$, the quasi energies $\mu$ and corresponding Fourier components of Floquet eigenstates can be readily calculated in a closed form from the hierarchical equations (A2) and (A3).\\ 
The first quasi energy is given by $\mu_1=-\omega_0/2$, and the Fourier components of the corresponding Floquet eigenstate $\mathbf{W}_{1}(t)$ read
\begin{eqnarray}
A_n & = \left\{  
\begin{array}{cc}
\mathcal{N}_1 & n=0 \\
\frac{\epsilon V_2^2}{n(-\omega_0+n\epsilon - \epsilon)}A_{n-2} & n=2,4,6,... \\
0 & {\rm otherwise}
\end{array}
\right.
\end{eqnarray}
\begin{eqnarray}
B_n & = \left\{  
\begin{array}{cc}
\frac{n+1}{V_2}A_{n+1} & n=1,3,5,... \\
0 & {\rm otherwise}
\end{array}
\right.
\end{eqnarray}
where $\mathcal{N}_1$ is a normalization constant.\\
The second   quasi energy is given by $\mu_2=\omega_0 /2$ with corresponding Floquet eigenstate $\mathbf{W}_2(t)$ with Fourier coefficients given by
\begin{eqnarray}
B_n & = \left\{  
\begin{array}{cc}
\mathcal{N}_2 & n=0 \\
\frac{\epsilon V_2^2}{n(\omega_0+n\epsilon - \epsilon)}B_{n-2} & n=2,4,6,... \\
0 & {\rm otherwise}
\end{array}
\right.
\end{eqnarray}
\begin{eqnarray}
A_n & = \left\{  
\begin{array}{cc}
\frac{n+1}{V_2}B_{n+1} & n=1,3,5,... \\
0 & {\rm otherwise}
\end{array}
\right.
\end{eqnarray}
where $\mathcal{N}_2$ is a normalization constant. An inspection of Eqs.(B3) and (B4) clearly shows that $\mathbf{W}_2(t)$ is level-2 dominant for a small value of $\epsilon$, with $\mathbf{W}_2(t) \simeq (0,1)^T+O( \epsilon)$. Similarly, from Eqs.(B1) and (B2) it follows that $\mathbf{W}_1(t)$ is level-1 dominant, i.e. 
$\mathbf{W}_1(t) \simeq (1,0)^T+O( \epsilon)$,  {\it provided that} $\omega_0$ is sufficiently far from $\epsilon(n-1)$ for any $n=2,4,6,...$. In fact, as $\epsilon$ approaches an odd resonance, let us say $\omega_0 \simeq (2N-1) \epsilon$, the denominator in the fraction on the right hand side of Eq.(B1) becomes extremely large (singular) for $n=2N$, so that the Fourier amplitudes $A_{2N}$, $B_{2N-1}$ become the dominant terms in the Fourier series. To avoid the singularity, the constant $\mathcal{N}_1$ should assume an extremely small value. Taking into account that 
 \begin{equation}
 \frac{B_{2N-1}}{A_{2N}} \simeq \frac{\omega_0}{\epsilon V_2} \sim 1/O(\epsilon).
  \end{equation}
  it follows that the dominant Fourier coefficient of $\mathbf{W}_1(t)$ near an odd resonance is $B_{2N-1}$, i.e. $\mathbf{W}_1(t)$ becomes level-2 dominant (like $\mathbf{W}_2$). Moreover, it can be readily shown that close to an odd resonance the two linearly independent solutions to Eqs.(12) and (13), namely $\mathbf{W}_1(t) \exp(i \omega_0 t/2)$ and $\mathbf{W}_2(t) \exp(-i \omega_0 t/2)$, become equal (parallel) each other and level-2 dominant. This is a clear signature that $\epsilon=\epsilon_N=\omega_0/(2N-1)$ is a Floquet exceptional point, i.e. a coalescence of {\it both} quasi energies and corresponding Floquet eigenstates occurs. In terms of the $2 \times 2$ Floquet matrix $\mathcal{R}$ entering in Eq.(15), this means that the eigenvalues and corresponding eigenvectors of $\mathcal{R}$ coalesce, i.e. that the matrix $\mathcal{R}$ is defective.


\end{document}